# Momentum Matching for 2D-3D Heterogeneous Ohmic van der Waals Contact

Tara Jabegu,[1#] Ningxin Li,[1,2#] Aisha Okmi,[3] Ben Tipton,[1] Ivan Vlassiouk,[4] Kai Xiao,[4] Yao Yao,[5*] Sidong Lei[1*]

[1] Department of Physics and Astronomy, Georgia State University, Atlanta, GA 30303, United States

[2] School of Electrical and Computer Engineering, Georgia Institute of Technology, Atlanta, GA 30303, United States

[3] Department of Physics, College of Science, Jazan University, Jazan 45142, Kingdom of Saudi Arabia

[4] Center for Nanophase Materials Sciences, Oak Ridge National Laboratory, Oak Ridge, TN 37830, United States

[5] Department of Physics, Kennesaw State University, Kennesaw, GA 30062, United States

**Abstract**:

Construction of ohmic contact is a long-standing challenge encountered by two-dimensional (2D) device fabrication and integration. van der Waals contacts, as a new solution for 2D contact construction, can effectively eliminate issues, such as Fermi-level pining and formation of Schottky barrier. Nevertheless, current research primarily considers energy band alignment, while ignoring the transverse momentum conservation of charge carriers during the quantum tunneling across the van der Waals contacts. In this study, by comparing the IV characteristics and tunneling spectra of graphene-silicon tunneling junctions with various interfacial transverse momentum distribution, we demonstrate the importance of charge carrier momentum in constructing high-performance 2D contact. Further, by conditioning the van der Waals contacts and minimizing the momentum mismatch, we successfully enhanced the quantum tunneling current with more than three orders of magnitude and obtain ohmic-like contact. Our study provide and effective method for the construction of direction 2D-3D contact with low resistance and can potentially benefit the heterogeneous of integration of 2D materials in post-CMOS architectures.

Two dimensional (2D) materials, such as graphene and transition metal dichalcogenides, are attractive candidates for future microelectronics featuring three-dimensional (3D) and heterogeneous integration.[1, 2] A long-standing challenge in this field is the construction of ohmic contacts on 2D materials.[3, 4] Thus far, various methods, such as side contact,[5, 6] phase engineering,[7, 8] orbital hybridization,[9] etc. Compared with these approaches, direct van der Waals (vdWs) contacts that couple 2D and other materials through vdW gaps avoid complex fabrication processes and can potentially address problems such as Fermi-level pinning and the formation of Schottky barrier.[10, 11] More importantly, vdWs contacts can potentially allow us to direct couple 2D materials to 3D semiconductor surfaces without involving metal electrodes. This solution is very suitable for the back-end-of-line (BEOL) process,[12] which integrates another layer of 2D material-based microelectronics on established 3D architectures to further increase the device density and functionality.

Considering the existence of vdW gap, a quantum tunneling should dominate the charge carrier transport through the contact interface, and it requires conservation of energy and conservation of transverse momentum (i.e., the charge carrier momentum component in parallel to the interfacial plane). On the other hand, most of the current efforts still focus on the band matching on the contact interfaces, *i.e.*, the conservation of energy, while ignoring the importance of charge carrier momentum distribution. Therefore, it is interesting and important to investigate the corresponding effects in order to complement the principles for the design of 2D contacts.

In this study, by employing graphene-silicon contacts as a representative of 2D-3D hetero-junctions, we investigated and confirmed the importance of momentum matching for the construction of ohmic vdW contacts. Further, thought interfacial conditioning, we compensated for the large transverse momentum mismatch between the p-type silicon Γ-point of and graphene K/K' points. The compensation leads to ohmic-like contacts with current level enhanced for more than three orders of magnitude. Our study indicates the momentum matching is another critical consideration for the fabrication of 2D contact, meanwhile, it provides a feasible approach for the construction of 2D-3D heterogenous integration.

**Effects of transverse momenta mismatch**

We employ junctions composed of graphene and heavily doped silicon (with a resistivity of 0.001~0.003Ω·cm) as examples to investigate the impact of charge carrier momenta during the quantum tunneling across the vdW contacts. Fig. 1a illustrates the cross-section of tunneling junctions employed in our study, and Fig. 1b shows the device fabrication workflow. We first perform photolithography and use buffered hydrogen fluoride (HF) acid to selectively etch $Si/SiO_2$ wafers and obtain round Si contact areas. The contact areas are passivated with Si-H bonds and directly in touch with graphene layers, while the other areas are covered with insulating $SiO_2$ films. Immediately after the etching, we transfer a graphene layer to the treated wafer and form Si-Gr contacts before the passivated Si contact areas get oxidized. The graphene transfer follows the procedure established in our earlier study [13] to obtain ultra-flat and clean interfaces. After the transfer, the devices are annealed at 300 °C under 200 mtorr $H_2/N_2$ environment for 60 min. Following this, a regular photolithography process and Argon-plasma etching are employed to pattern the continuous graphene films into isolated disks with a diameter larger than the round Si contact areas. Then, a pair of bracket-shaped electrodes are fabricated on each of the graphene disks and form side contacts with them. The gold electrodes do not overlap with Si contact areas. Refer to the **Experimental Method** section for more details. Fig. 1c shows the optical image of one device employed in our study.

We select and design the Si-Gr junctions for our study based on the following considerations. First, silicon is the most widely used 3D semiconductor in microelectronics, while graphene has a six-fold K/K' valley band structure configuration similar to many other 2D materials, such as $MoS_2$. Therefore, their combination can serve as a good representative with significant versatility. Also, graphene is a zero-band gap material so that the energy band matching (energy conservation) condition is always met, allowing us to focus on the momentum effects. Further, by choosing n-type and p-type silicon, we can obtain and compare two momenta distributions on the (100)-silicon-graphene interfaces to examine their corresponding effects, as illustrated in Fig. 2a. (Here, we only need to consider the momentum components in the junction interfacial plan, *i.e.*, the transverse momenta that is required to be conserved during the quantum tunneling, whereas the longitudinal components do not experience this restriction.) Because graphene has a 2D structure, its Fermi surface consists of six one-dimensional rings. Thus, we call them Fermi rings hereafter.

Both types of silicon wafers are heavily doped with a resistivity of 0.001~0.005 $\Omega \cdot cm$. Due to the high doping level and graphene quantum capacitance effect,[14] we can ignore the formation and thickness of depletion region in silicon and its effect on tunneling barrier. This assumption is confirmed by the temperature dependent IV measurement. Fig. 2b shows the current levels of the p-Si-Gr and n-Si-Gr junctions with a +0.5 V bias under different temperatures. (Fig. S1 shows the original IV curves.) No obvious change in current levels is observed in either configuration, indicating quantum tunneling dominates the interfacial charge transport.

We further compared the IV curves collected from respective devices at 5 K, as shown in Fig. 2c. Although both types of silicon wafers have the same conductivity, the n-Si-Gr clearly shows a tunneling current with two orders of magnitude higher than the p-counterpart. Two possible mechanisms can potentially explain this observation. First, n-Si-Gr and p-Si-Gr junctions have different band structure alignments on the interfaces and cause variation in tunneling current levels. The other possibility is the n-Si-Gr junction has a smaller transverse momentum mismatch (from X-valleys to K/K' valleys, as shown in Fig. 1a) that the p-counterpart (from Γ-point to K/K' valleys), hence, better meeting the requirement of momentum conservation and having a larger tunneling current. To determine the dominating reason, we further performed tunneling spectroscopy (dI/dV) measurements on these two devices. Fig. 2d illustrates the schematic of the experimental setup for the spectroscopic measurements, and Fig. 2e shows the normalized tunneling spectra and reveals several interesting characters. The n-Si-Gr junction clearly exhibits a V-shape feature with the Dirac point locating at 70 mV, indicating a p-type doping in graphene layer. In contrast, the p-Si-Gr junction shows no similar feature, but has a very flat curve between -0.4 to +0.4 V. These observations suggest that momentum instead of energy match is the main cause for the current level difference because of the following reasons.

First, the change of energy band alignment is not in consistency with the variant in current level. In both cases, since the graphene layer in the n-Si-Gr device is p-doped, the p-Si-Gr should also be p-doped with a Fermi level moving to a more positive position. Therefore, p-Si should increase the charge carrier density in the graphene layer and bring a higher current level. This is in contradictory with our IV experimental observations. Further, we can consider the flat tunneling spectroscopic curve observed on the p-Si-Gr device as a characteristics of the pseudo-gap induced by phonon-assisted inelastic tunneling.[15] Specifically, due to the large transverse momentum mismatch between p-Si and graphene, as illustrated in Fig. 2a, elastic tunneling is unlikely to occur. In this condition, if a spontaneous phonon emission can compensate for the momentum mismatch, meanwhile resonating with the energy difference between the electron emitter and accepter states, an inelastic tunneling takes place, as illustrated in Fig. 2e inset.

Considering both aspects, we can construct the picture that due to the large transverse momentum mismatch, p-Si-Gr junction has a much lower tunneling current. When a higher bias voltage is applied, phonon-assisted inelastic tunneling takes place and provides the necessary transverse momentum. This explains the observed zero-bias gap and the much lower current level. In contrast, the smaller momentum mismatch on the n-Si-Gr junction leads to a much more effective quantum tunneling and better contact.

On the other hand, we still need more experimental evidence to support the above picture. If we can find a way to compensate for the large momentum mismatch in the p-Si-Gr junction and verify that it can effectively enhance the tunneling current, the picture can be properly justified. Further, both p-type and n-type silicon wafers are widely used for practical applications in microelectronics. It is irrational to restrict the device design to one configuration over the other. Thus, it is necessary to find a solution to build high quality p-Si-Gr contact. More importantly, the verified picture, in turn, can potentially complement our understanding on vdW contacts, particularly, the importance of momentum match. In this regard, we design additional experiments as discussed as follows.

**Interfacial conditioning for ohmic-like contacts**

In order to compensate for the momentum mismatch between p-Si and graphene, we introduced a thin layer of gold nanoparticle on the interfaces with two significances. First, it has a work function of 5.38 eV [16] that is very close to the energy level of silicon valence band top, which is 5.17 eV and equals to electron affinities of 4.05 eV[17] [need more references, better from database] plus the band gap of 1.12 eV.[18] Thus, the introduction of Au nanoparticles will not significantly change the Fermi level in p-Si. Second, gold has a very large Fermi radius, so that we expect it can scatter electrons and compensate for the momentum mismatch to facilitate the quantum tunneling, as illustrated in Fig. 3a. Experimentally, we thermally deposited 2 nm of gold on the HF-treated Si surfaces before transferring the graphene layers. The deposition is controlled by a regular crystal thickness monitor. Since the gold thickness is very thin, it does not form a continuous and conductive film. Instead, numerous isolated gold nanoparticles scatter across the silicon surface, as shown in Fig. 3b. Fig. 3c compares the IV curves of p-Si-Gr junction with and without Au-nanoparticles and clearly exhibits a tunneling current enhancement with three orders of magnitude. IV curve is almost linear, indicating an ohmic-like contact.

In order to more insightfully inspect the interfacial transport process, we also captured the tunneling spectra, as shown in Fig. 3d. Compared with the device without gold conditioning (Fig. 2e blue curve), a narrow and sharp gap with a width of about 12 mV appears. As the temperature varies, the width of the gap does not change significantly. The $2^{nd}$-order spectra shown in Fig. R2 better define the gap and its temperature- invariance. We cannot assign this gap to any of the energy band structures of graphene, gold, or silicon. Instead, transverse momentum mismatch and phonon-assisted tunneling explains the [19]phenomenon.

Specifically, the interfacial transport has two steps with the introduction of Au-nanoparticles: electron propagation between gold and silicon, as well as quantum tunneling between graphene and gold. Because the gold work function is very close to silicon ionization energy, an ohmic contact forms between them, as shown in Fig. R3. Also, we did not observe a similar gap feature on the Si-Au junction, indicating this is character of the Gr-Au contact. Therefore, the gap feature cannot be attributed to the Si-Au contact.

Instead, it must come from the Au-Gr tunneling. After a closer inspection, we can find that although gold has a very large Fermi surface radius of about 1.36 Å$^{-1}$, (which is calculated from Ref.). Thus, there is still a 0.34 Å$^{-1}$ momentum mismatch to graphene K/K' points (1.70 Å$^{-1}$ away from the $\Gamma$ point) and it requires assistance of phonon scattering for tunneling, as denoted by the red arrow in Fig. 3a. Considering that the gap has a 12mV width, the phonon energy should be 6meV, *i.e.*, half of the gap width. By investigating the phonon dispersion relations of graphene and gold, we find graphene out-of-plane acoustic (ZA) phonon[20, 21] can contribute to the phonon-assisted inelastic tunneling, in consistent with early scanning tunneling microscopy investigation on graphene.[15] Furthermore, the tunneling spectra do not experience significant change in phonon peak position over a broad temperature range, suggesting that the graphene ZA phonon or gold phonon dominate the inelastic tunneling. This is because this phonon mode is the only one with a relatively large momentum and a low energy that facilitates the Au-Gr tunneling. IV measurement (Fig. R4) shows very weak temperature-dependence, suggesting a spontaneous phonon emission process dominates the inelastic tunneling through the entire temperature range.

To better justify the above explanation and exclude other possible mechanisms, particularly modification in band structure alignment that can possibly result in better contact, we also deposit the Au nanoparticles on the exposed graphene surfaces, instead of the Si-Gr interfaces. In principle, both configurations should deliver the same band alignment after reaching thermal equilibrium. However, the decoration of Au nanoparticles on the graphene surface does not bring current level enhancement to the same level as the devices with gold on the interface. This indicates that band alignment is not the primary reason for the Au-induced tunneling current enhancement.

Despite that, we can still observe about 25× current enhancement after the Au-decoration on the surface, as shown in Fig. 3c inset on the lower right conner. More interesting, we find a similar but wider gap-feature in the tunneling spectra, as shown in Fig. 3e. The gap width does not vary as the temperature changes, either. (The 2$^{nd}$-order spectra shown in Fig. R5 can better demonstrate this invariance.) A two-step tunneling model illustrated in Fig. 3f can explain this phenomenon. Due to the large momentum mismatch between p-Si and graphene, the tunneling electrons have very limited interaction with the graphene layer. Thus, the graphene layer is nearly transparent to the electrons, allowing them to reach the Au particles. Inside the particles, electrons experience the elastic scattering, followed by phonon-assisted inelastic scattering, and then tunnel back to the graphene layer. Because the tunneling process consists of two steps and the first one involves a much wider tunneling barrier, the current level is much lower than the one given by the device with interfacial Au-particles. On the other hand, the phonon-assisted inelastic tunneling still exists in the second step, and thus, resulting in a much higher current level than the device with no gold decoration. The existence of the phonon-assisted tunneling is also supported by the gap feature in the tunneling spectra.

Future, this two-step tunneling picture also explains the broader gap appearing the tunneling spectra. Since the tunneling electrons first reach the Au-particles, they can push their electrical potential to the more negative direction, creating potential drop between gold and graphene to trigger the second step. Therefore, we can employ an in-series resistors model to simplify the process, as illustrated in Fig. 3f. The resistance between silicon and gold acts as a voltage divider that reduces the effective bias applied between gold and graphene. Therefore, a high voltage should be applied on the entire junction to induce the photon scattering, resulting in a larger gap in the spectra.

The above experimental observations further confirmed that the gold-induced current enhancement is not a result of better band structure alignment. The primary function of gold is not a metal electrode connecting silicon and graphene, because Au-particles on graphene surface can also enhance the current and bring a similar gap feature in the tunneling spectra, although the voltage dividing effect makes the inelastic scattering less effective. Considering all these factors, we can interpret the current enhancement as a synergic effect of two scattering processes. Gold, with a large Fermi radius, serves as an elastic scattering medium and provides electrons with a major part of the transverse momentum, as denoted by the purple arrow in Fig. 3a. Meanwhile, the low energy phonons inelastic scattering (the red arrow) process compensates for the rest minor momentum mismatch and facilitates the quantum tunneling.

**Effect of gold deposition thickness and position**

This above model does not play any requirement on the continuity or conductivity of gold. As mentioned earlier, we deposit 2 nm of gold that tends to form isolated nanoparticles instead of continuous film. Thus, gold is considered a scattering medium instead of electrode. To further verify this picture, we also fabricated p-Si-Gr junctions different gold deposition thicknesses in two extreme scenarios as 3 Å and 10 nm. Figs. 4a and 4b show the corresponding IV and tunneling spectra. Fig. 4a inset also compares the tunneling current level under the same sample bias.

For a 10 nm gold deposition, a continuous gold film forms, but the current level is slightly smaller than that with 2 nm gold deposition. The tunneling spectrum sustains a similar characteristic. Thus, a thicker and continuous gold layer does not bring additional benefits for the quantum tunneling process. On the 2 Å sample, we can still observe notable current enhancement, although it is not as strong as the 2 nm case. The IV curve shows slight non-linearity. These facts indicate that the Si-Gr junction conductivity is very sensitive to gold decoration. As long as there is a trace-amount of gold, can serve as scattering centers to minimize the momentum mismatch, the quantum tunneling channel opens with a current increasing for several order of magnitude. On the other hand, sufficient gold is still needed to provide enough transport channels for high interfacial conductivity.

Although the current enhancement suggests a strong gold scattering in the 2 Å sample, the tunneling spectrum does not have a well-defined narrow zero-bias gap, as shown in Fig. 4b, which is distinct from 2 nm and 10 nm cases. We can explain it in such a way that when the gold deposition is very thin or the particle (cluster) size is very small, the quantum uncertainty relationship ($\Delta x \cdot \Delta k \sim 1$) starts to play a role. The 3 Å gold size corresponding to a 0.33 Å$^{-1}$ wavevector, which explains the disappearance of the phonon gap. However, since the deposition is so thin with much fewer scattering centers, the current level is lower.

**In summary**, through the study on Si-Gr heterojunctions, we demonstrate the momentum distribution and transverse momentum match is another critical consideration besides energy band alignment for the construction of ohmic contacts on 2D materials. Meanwhile, our study also presents an effective approach for the construction of vdW contact in the scenario that transverse momentum match cannot be intrinsically met. By introducing elastic scattering media, such as gold, we successfully minimize the momentum mismatch between p-type silicon and graphene and obtain a contact current enhancement of more than three orders of magnitude. Our study completes the principles and guidelines for the construction of high-performance 2D contacts, and provides feasible approaches for the development of 2D-3D heterogeneous integrated microelectronics.

**Experimental Methods**

CVD Growth of Graphene: A homebuilt CVD system had been employed to grow monolayer graphene. A 0.025 mm thick copper foil (CU000358 from Goodfellow) was electropolished in 80% phosphoric acid ($H_3PO_4$) at 8V DC power supply and the copper foil was dried with $N_2$ gun then immediately inserted inside a quartz tube of CVD system and sealed. It was annealed for 20min at 1040ºC with 6sccm supply of argon and hydrogen (Ar 90%: $H_2$ 10%) then the growth was carried out at 1040° C by controlling methane ($CH_4$) flowrate at 35 sccm and a mixture of argon and hydrogen (Ar 90%: $H_2$ 10%) at 6 sccm for 30min.

Etching: Two step etching process was followed to get free floating monolayer graphene. First, the bottom side of the copper foil was etched to remove unwanted graphene using RF argon plasma. Secondly, the copper foil was etched slowly in 0.1M ammonium persulfate(($NH_4)_2S_2O_8$) (from Sigma-Aldrich) solution in enclosed $N_2$ environment to prevent oxidation. This etching solution was then replaced by DI water 3 times making the free floating graphene ready for the transfer to a substrate.

Device Fabrication: Free floating graphene on DI water was transferred using polymer free wet transfer to a special silicon substrate having $SiO_2$ layer with disc shape openings of varying sizes followed by annealing at 300C for 60 min with 35sccm of a mixture of argon and hydrogen (Ar 90%: $H_2$ 10%) at 200

mTorr vacuum. Then, following multiple photolithography process, the graphene was etched using RF oxygen plasma leaving discs of graphene covering the $SiO_2$ openings. Two gold brackets were fabricated around the circumference of the graphene discs as electrodes by thermal deposition of 5nm Cr and 45nm Au followed by lift off in acetone solution. A certain thickness of Au was deposited on the substrate just before the graphene transfer or after the electrode bracket fabrication depending on the requirements.

Characterization: The IV and tunneling spectrum measurements were done on a homebuilt cryogenic probe station. IV measurement was done with a source meter unit (SMU, Keithley 2450). The first order tunneling spectrum measurement was done with a digital lock-in amplifier (SR 830 DSP Lock-in Amplifier) and the second order tunneling spectrum was obtained with an analog lock-in amplifier (SR2124 Dual-Phase Analog Lock-In Amplifier). For both spectrum measurements a modulated DC bias was supplied (Keithley 2450 and SRS DS360 Function Generator). AFM scanning of graphene was performed on the Veeco MultiMode AFM system under tapping mode and the SEM was performed on a TESCAN VEGA3 system.

**Acknowledgement**

This work is supported by National Science Foundation DMR-2105126 and ECCS-2238564.

**Figures:**

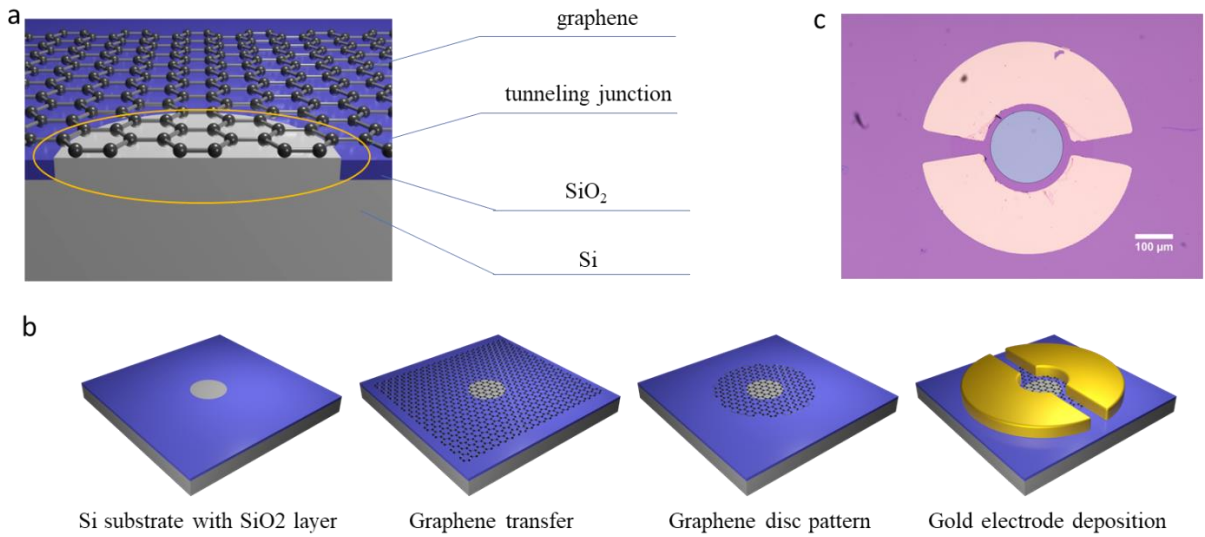

**Figure 1**. (a) Schematic of quantum tunneling from Silicon to graphene. (b) Schematic representation of workflow of device fabrication. (c) Optical image of Silicon-Graphene tunneling junction device.

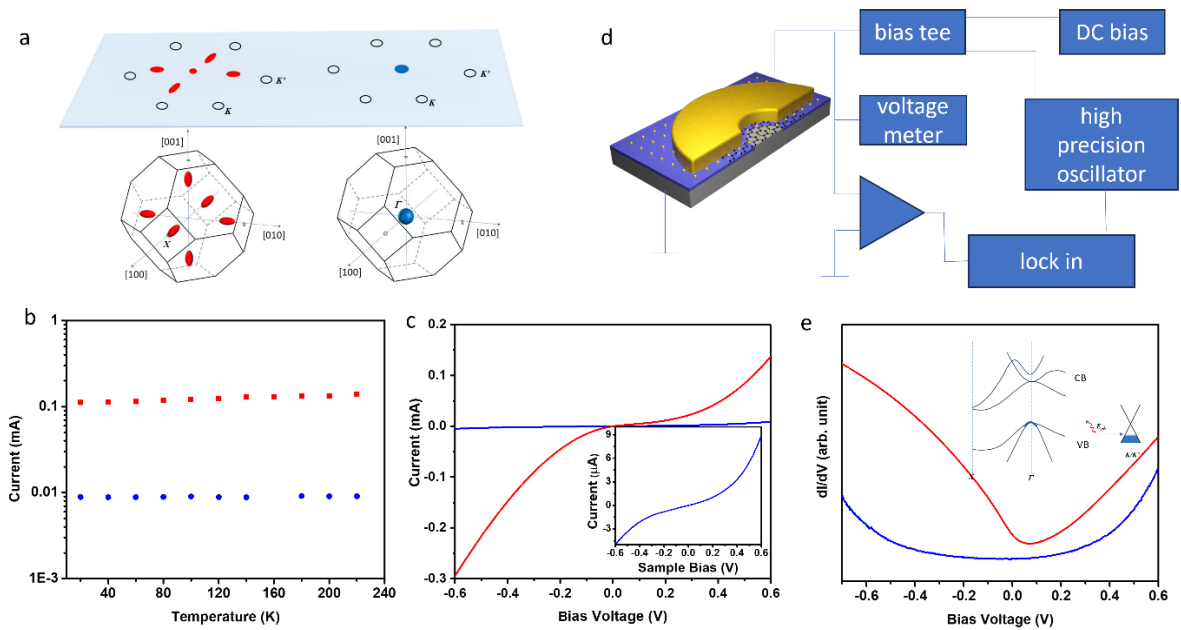

**Figure 2.** (a) Fermi surface diagram of Silicon and graphene showing momenta mismatch (n-type silicon on left and p-type silicon on right). (b) Current through the junction at 0.5V bias at different temperature. (c) IV curves of p-type and n-type silicon devices with magnified IV curve of p-type device in inset. (d) Spectrum measurement schematic. (e) First order spectrum graphs of p-type and n-type silicon devices with energy level diagrams in inset.

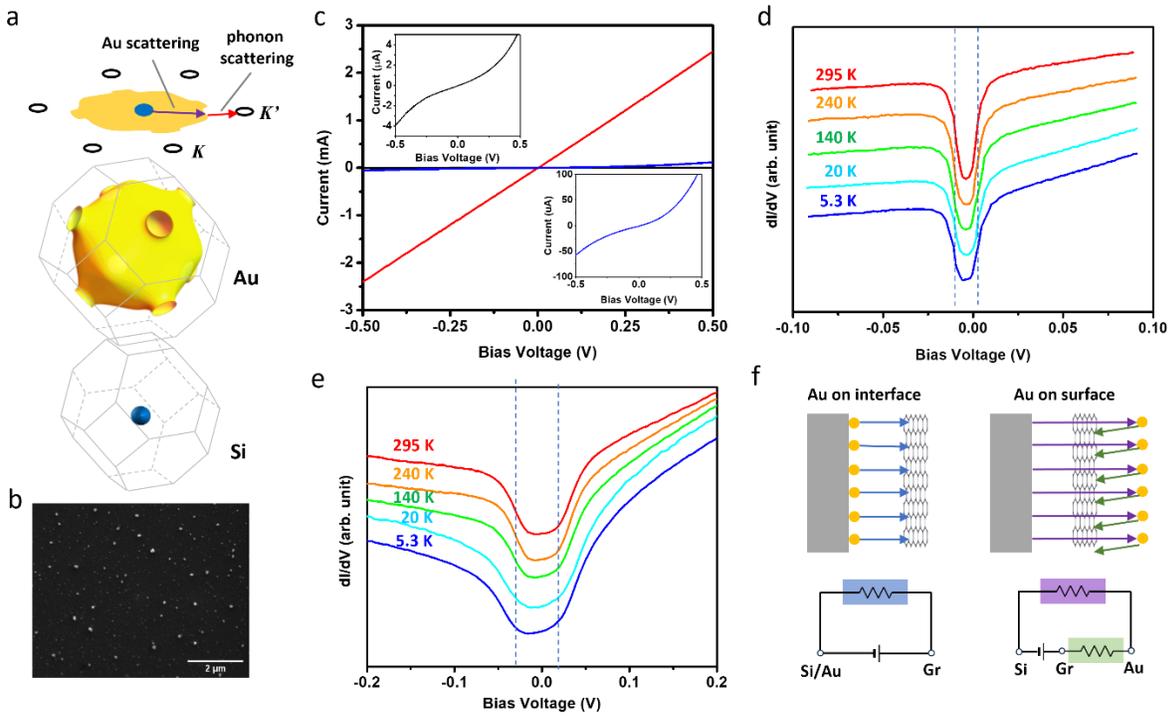

**Figure 3.** (a) Fermi surface diagram of p-type silicon and graphene with gold introduced in the middle. (b) SEM image of 2nm gold deposition on silicon. (c) IV curves of device without gold in the middle and device with 2nm gold in the middle (inset upper left is the IV curve magnified of device without gold and the lower right is the IV curve of device with 2nm gold on top of graphene). (d) First order spectrum of p-type device with 2nm gold. (e) First order spectrum of p-type device with 2nm gold on top. (f) Tunneling models of devices with gold in middle and gold on top.

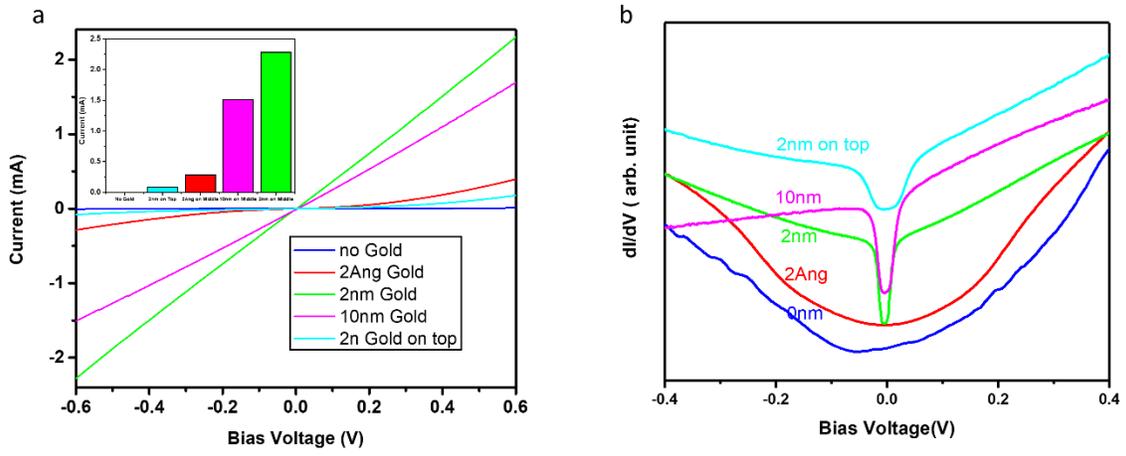

**Figure 4.** (a) IV curves of p-type silicon device with different gold deposition thickness and surface and inset is the current levels at 0.6V bias. (b) First order tunneling spectrums of p-type devices with different gold deposition thickness and surface.

**Supporting Figures:**

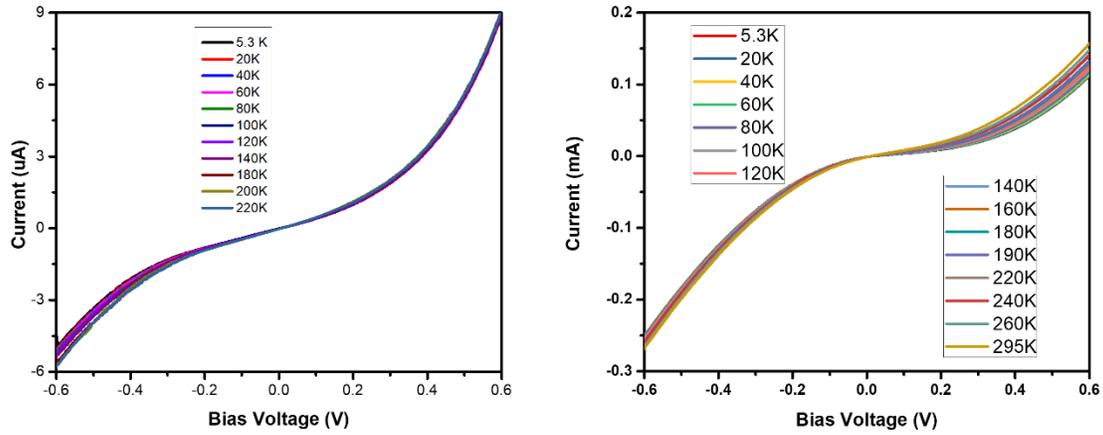

**Figure S1.** IV measurements of p-type silicon device without gold on left and with 2nm gold in the middle on right at different temperature.

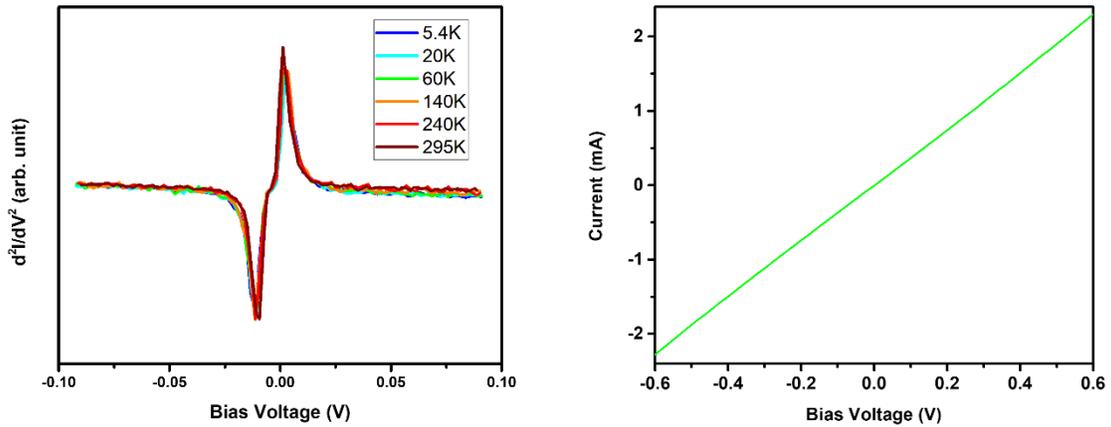
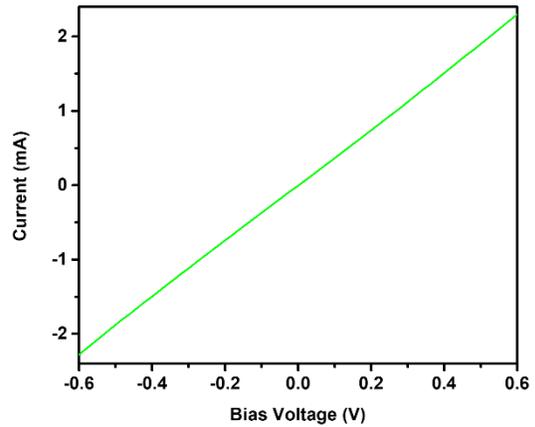

**Figure S2.** Second order tunneling spectrum of p-type device with 2nm gold deposition in between silicon and graphene.

**Figure S3.** IV measurement of p-type silicon device with 2nm gold deposition in between silicon and graphene showing ohmic like contact.

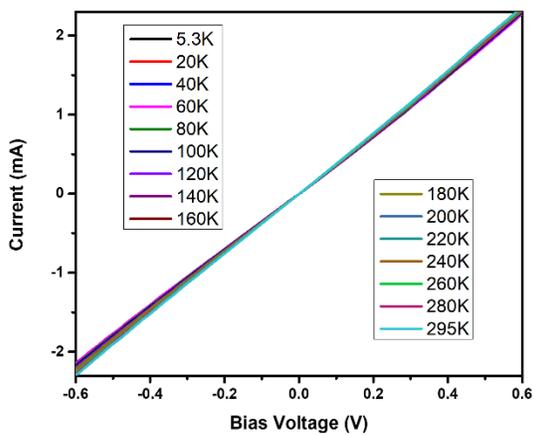 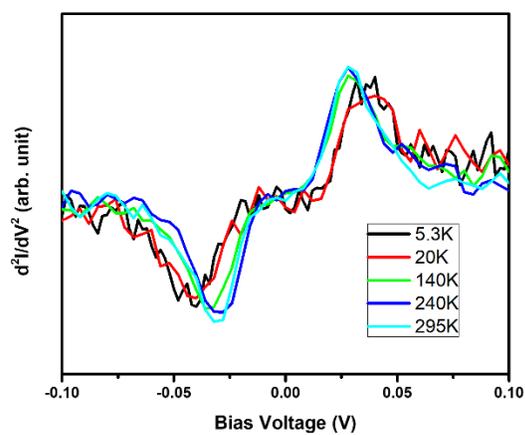

**Figure S4.** IV curves of p-type device with 2nm gold deposition in between silicon and graphene at different temperature.

**Figure S5.** Second order tunneling spectrum of p-type device with 2nm gold deposition on top of graphene at different temperature.